\documentstyle[epsfig]{aipproc}
\def\be{\begin{equation}}
\def\ee{\end{equation}}
\begin{document}
\title{The Time-Dependent Local Density Approximation for
Collective Excitations of Atomic Clusters}
\author{G.F. Bertsch}
\address{Physics Department and Institute for Nuclear
Theory\\
University of Washington, Seattle, WA 98195 USA\\
E-mail: bertsch@phys.washington.edu\\
and\\}
\author{K. Yabana} 
\address{Department of Physics, Niigata University\\
Niigata, Japan\\
E-mail:yabana@carrot.sc.niigata-u.ac.jp\\
}
\maketitle
\def\C60{C$_{60}$}
\def\eq#1{eq. (\ref{#1})}
\begin{abstract}
We discuss the calculation of collective excitations in atomic
clusters using the time-dependent local density approximation.
In principle there are many formulations of the TDLDA, but we
have found that a particularly efficient method for large
clusters is to use a 
coordinate
space mesh and the algorithms for the operators and the evolution
equations that had been developed for the nuclear time-dependent Hartree-Fock
theory.  The TDLDA works remarkably well to describe the strong
excitations in alkali metal clusters and in carbon clusters.  We
show as an example the benzene molecule, which has two
strong features in its spectrum.  The systematics of the linear
carbon chains is well reproduced, and may be understood in
rather simple terms.
\end{abstract}

\section{Introduction}  
The time-dependent local density approximation (TDLDA) is a
powerful tool to calculate the quantum mechanical motion
of electrons in condensed matter systems.  In my talk
I want to describe to you the numerical techniques we
use \cite{ya96}, which were borrowed from nuclear physics\cite{fl78}.  I then
want to show you a survey of some of the results, 
concluding with new work on the behavior of electrons in 
elongated chains.  The
time-independent LDA is known as Density Functional Theory
and is well-established in condensed matter physics \cite{jo89} as
a predictive {\it ab initio} theory that is practical for structures
beyond the range of quantum chemistry methods [4-7].
The equations of the LDA are easy to write down.  The energy
function is given by the expression
\be
E= \int d^3 r \Biggl[\sum_i{|\nabla \phi_i|^2\over 2 m} +
+\sum_I V_{ion}(r-r_I) \rho(r) +
\ee
$$ + {1\over 2} \int d^3 r'{e^2 \over |r-r'|}
\rho(r)\rho(r') + v_{ex}\bigl(\rho(r)\bigr)\Biggr]
$$
where the $\phi_i$ are the single-particle wave functions and
$\rho(r)=\sum_i|\phi_i(r)|^2$ is the density.  The potentials
included in the energy function are $V_{ion}$, the ionic potential,
$e^2/|r-r'|$, the direct Coulomb interaction between electrons, 
and $v_{ex}(\rho)$, the local-density approximation to the
exchange-correlation energy.
The static LDA theory
is obtained by minimizing this energy function, requiring only that
the $\phi_i$ be orthonormal.  That gives
the Kohn-Sham equations for the wave functions
$ \phi_i$.  The TDLDA equations are very similar, with the
``energy" (actually the Lagrange multiplier) in the Kohn-Sham
equation replaced by the time derivative $i\hbar\partial_t$,
\be
i \hbar \partial_t\phi_i(\vec r,t)
= H \phi_i(\vec r,t).
\ee
with
$$
H= -\frac{\hbar^2}{2m}\nabla^2 
         + \sum_{I} V_{ion}(\vec r - \vec R_I)
         + e^2 \int d\vec r' \frac{\rho(\vec r',t)}{|\vec r - \vec r'|}
         + {d v_{xc} \over d \rho}\Big|_{r,t}.
$$
This is almost all I want to say in general.  One always uses
pseudopotentials for the ions\cite{tr91,kl82}, to eliminate the
core electrons with their large energy scales. The LDA has well-known
difficulties in describing the electronic excitations of systems,
attributable to the simplified treatment of exchange.  However, the
collective motion is rather insensitive to the nonlocality of the
exchange, and the TDLDA is much more reliable than one would
expect from looking at energy gaps.

\section{Numerical}

For most applications it is sufficient to consider small 
deviations from the static solution, and then there are
a number of approaches to solve the small-amplitude
TDLDA equations.  Nuclear physicists are most familiar
with the configuration representation, which I'll just
mention briefly.  Here one
represents the time-dependent wave function in the 
basis of the static solutions,
\be
\phi_i(t) = \phi_i(0) + \sum_j a_{ij}(t) \phi_j(0).
\ee
The equation of motion is Fourier transformed to frequency,
and the matrix equations satisfied by the particle-hole
amplitudes 
$\tilde a_{ij}(\omega)$ are the RPA equations of motion.  This
method becomes inefficient for large systems, because the
the dimensionality of the matrix grows as the product of the
number of particle and hole orbitals.  In the more efficient
methods, the dimensionality increases linearly with the number of 
particles.  One method that is very
popular is to calculate the density-density response function.
This is the density change induced at a point $r$ by a 
density perturbation at a point $r'$.  Because of the local
density approximation, one can calculate the full response
from the response of non-interacting electrons.  The number
of mesh points to represent the response increases as
the volume of the system, and thus only linearly with 
the number of particles,
making it more efficient for large systems.
The method has been used extensively for systems with
spherical symmetry\cite{ek84,be75}; recently it has been applied also to
clusters in three dimensions\cite{ru96}.

We use another method whose dimensionality also scales linearly
with the system size.  This is the straightforward solution
of the time-dependent equations of motion.  The number
of variables is the number of points at which wave
function amplitudes are represented.  This can be of
the order of 50,000 for some of the examples we consider,
so some care must be taken to use efficient numerical
algorithms.  We have simply adopted the technique of
ref. \cite{fl78} who investigated the TDHF theory of
nuclear dynamics.

We use a uniform mesh in coordinate space to represent
the wave function.  The shape of the gridded volume 
can be spherical, cylindrical, or rectangular, depending
on the cluster under consideration.  
If the equations were linear in $\phi$, they could efficiently
integrated by using the Taylor series expansion of the
evolution operator, $\phi(t) = \exp(-i H t) \phi = 1+t H\phi +
t^2 H^2\phi/2+...$  In practice we can ignore the change
of H dependence
of $H$ on $\phi$ for some small time interval $\tau$, and
use the Taylor series integrator over that interval.  We typically
use fourth order:
\be
e^{-iH\tau}\phi = \sum_{n=0}^4  {(i\tau)^n\over n!}H^n\phi
\ee
Considerable care must be given to defining the $H$ to 
propagate in the above equation.  An important consideration
is that the algorithm conserve energy to a very high accuracy.  
This can be assured with an implicit definition of $H$ in
terms of the densities $\rho(t)$ and $\rho(t+\tau)$ at the
beginning and end of the interval $\tau$.  The definition
is
\be
H \phi(r) = \Big[ {-\nabla^2\over 2m} +
{v\big(\rho(r,t+\tau)\bigr) -v\big(\rho(r,t)\bigr)\over
\rho(r,t+\tau) - \rho(r,t)}\Bigr] \phi(r).
\ee
In practice we do not calculate this exactly but rather 
estimate the Hamiltonian at $t+\tau/2$ using a 
predictor-corrector method.  In our algorithm,
first the potential is extrapolated forward by half a time step
with a 3-point difference formula to get a trial potential
$V_{trial}(t+\tau/2)$. The time integration is then performed
using $V_{trial}$
to get a trial density
$\rho_{trial}(t+\tau)$,
and the density at time $t+\tau/2$ is estimated as $\rho(t+\tau/2)=
(\rho(t)+\rho_{trial})/2$.  The potential obtained with this density
is used to once more 
integrate the wave function from $t$ to $t+\tau$.

To generate the response of the system, we perturb the
ground state wave function and measure the response as
a function of time.
The initial wave functions are taken as
$
\phi_i(\vec r,0) = {\rm e}^{ikz}\phi_i^o(\vec r)
$
in which $\phi_i^o$ represent the static Kohn-Sham solutions. 
We are interested in the optical response, requiring the
perturbing parameter $k$ to is small.  This ensures
also that the response will be in the linear regime.
The real time evolution of the dipole moment is obtained as
$
z(t) = \sum_i <\phi_i(t) | z | \phi_i(t)>,
$
and its Fourier transform in time gives the dipole strength
function.

Summarizing, our algorithm to integrate the TDLDA equations
has as numerical parameters the spatial mesh size, 
$\Delta x$, the number of mesh points $M$, the time step $\tau$, and 
the total length of time integration $T$.  The actual values
for the different systems will be given below.

\section{Physics}
   Our interest is in the electronic excitations, and we
will leave aside completely the problem of determining the
ionic geometries of the clusters.  In classical electromagnetic
theory, the dipole strength in spherical metal clusters is concentrated in
the Mie resonance.  The resonance may be sharp or not depending
on the cluster, and it may be present also in clusters of
different shapes.
\subsection{Alkali metals}
 The simple, sharp Mie resonance behavior
is reproduced by jellium model of alkali metal clusters\cite{ek84},
which however neglect the perturbing effects
of the ionic cores. It is therefore interesting to see to what
extent the ionic cores affect the physics.  I will first consider
a system where the jellium approximation is remarkably good.
This is the for clusters of sodium atoms.  An example of
a calculation is the sodium cluster Na$_{147}$.  This was
chosen because 147 atoms can be placed in a icosahedral
structure, which is as close a one can get to spherical 
structure with discrete atoms.  The numerical parameters of
the calculation are: $\Delta x = 0.8$ \AA, $\tau/\hbar=0.01$ eV$^{-1}$,
$M=28000$ and $N_t=5000$.  Because the valence electron is
in an $s$-wave, and the short-distance part of the wave function
has been smoothed away by the pseudopotential, a rather coarse
$\Delta x$ is sufficient.  The total time integration is
$\tau N_t/\hbar= 50$ eV$^{-1}$ which in principle allows 
structures on the scale of 0.02 eV to be resolved.  Fig. 1
shows the response, compared with the spherical jellium model.
Both descriptions show a sharp peak at just over 3 eV excitation,
and some background extending in both directions. The two
theories are quite close.  This shows that the pseudopotential
in sodium has little effect.  Experimentally, sodium clusters
show a fairly sharp resonance, but the energy is lower by
about 15\%. 

\begin{figure}[t!]
\centerline{\epsfig{file=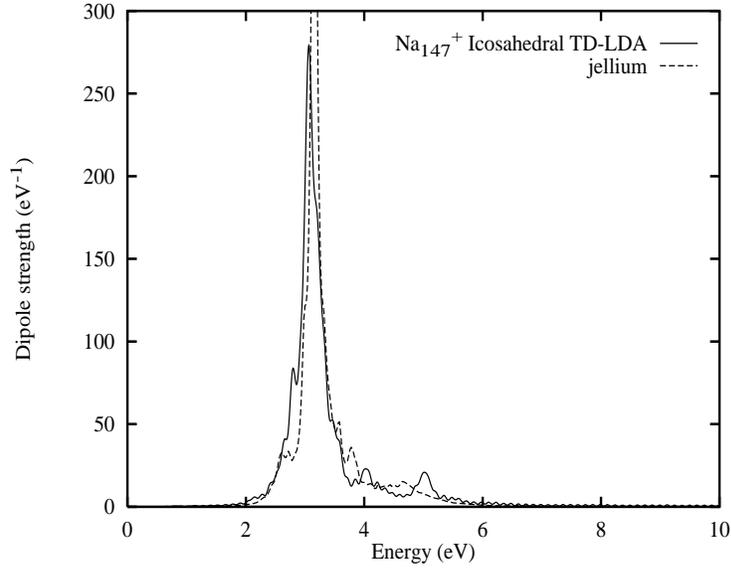,height=7.5cm,width=10cm}}
\caption{Dipole response of a sodium cluster, Na$_{147}^+$,
comparing the spherical jellium model with the full pseudopotential
TDLDA calculation.}
\vspace*{10pt}
\end{figure}

\begin{figure}[t!]
\centerline{\epsfig{file=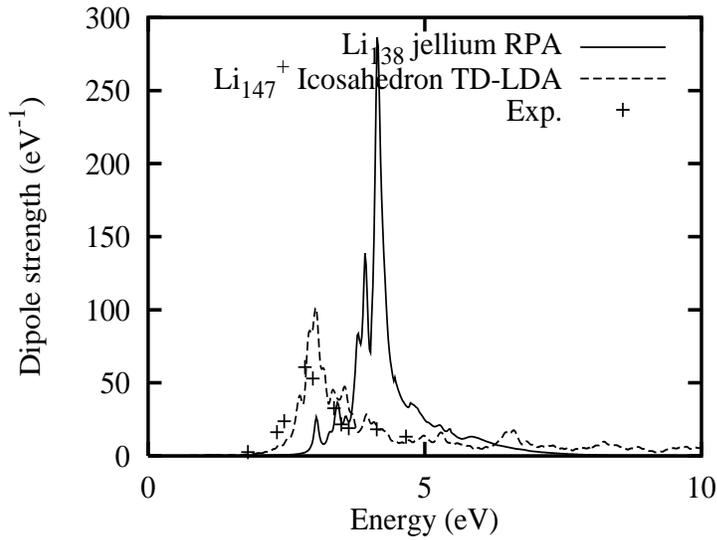,height=7.5cm,width=10cm}}
\caption{Dipole response of lithium clusters, comparing the jellium
model, the full TDLDA calculation, and experiment [13].}
\vspace*{10pt}
\end{figure}

The next case, lithium,  shows a more dramatic difference between 
the jellium model and the full pseudopotential TDLDA.  
Fig.~2 shows the response of Li$_{147}^+$,
comparing the full pseudopotential LDA with the jellium model.
The Mie resonance has been shifted down by the pseudopotential
interactions and broadened.  The reason for the difference
compared to sodium is not hard to find\cite{ya95}.
The scattering effect
of the pseudopotential is much larger in the case of Li, and
the scattering gives the electrons an effective mass larger than
one.  In the jellium model the ion behaves as a uniform positive
charge density filling space within the lattice.  The scattering
effect of the pseudopotential may be seen by examining the
scattering from the difference between the pseudopotential and
a uniform positive charge.  Thus, we are led to examine the
scattering from the difference potential,
$$   \Delta V = V_{i}(r) + {e^2\over
r_0}\Big({3\over2}-{1\over2}({r\over r_0})^2\Big)~~~~  r<r_s $$
$$ = V_{i}(r) +{e^2\over r} ~~~~r>r_s $$
When we do this we find that both the $s$- and $p$-wave scattering
phase shifts for sodium are small, showing that effects of the
pseudopotential are weak.  On the other hand, in the case of Li,
the $s$-potential is repulsive but the $p$-wave potential is 
attractive.  This implies that there will be a large backward
angle scattering, making the single-particle wave functions 
rather different than the jellium wave functions.  We believe
this aspect is responsible for the broadening of the Mie resonance
in Li.  Another consequence of the two potentials is that the
effective forward scattering potential will be momentum-dependent,
becoming more attractive as the momentum increases and the amount
of $p$-wave increases. This lowers the Mie resonance frequency.
\subsection{Carbon structures}
Carbon is more difficult because the $p$-valence electrons are
rather tightly bound and require a finer mesh.  In practice,
a mesh size of $\Delta x = 0.3 $ \AA~is adequate to calculate
the electronic response of carbon structures, given their geometry.
As a first test, we examine the response of the benzene molecule,
which has been well-studied up to 30 eV with synchrotron photon
sources\cite{ko72,hi91}.  For the calculation,
we used a pseudopotential for the hydrogen atoms as well as
the carbon atoms, and used the actual geometry of the molecule.
The grid of mesh mesh points was limited to points in a sphere
of radius 7~\AA.  This requires about 50,000 mesh points.  Finally,
because of the tighter binding of the $p$-waves, a time
step a factor of 10 smaller than that for the alkali metals 
was required, $\tau = 0.001$ eV$^{-1}$.  The resulting response
is plotted in Fig. 3, showing the Fourier transform of the real-time
response computed over a time interval of 30 $\hbar$/eV.  This is
compared to the strength function measured in ref. \cite{ko72},
which an adjusted normalization (which was arbitrary in ref. \cite{ko72}.
The main
features of the response are a sharp peak at 6.9 eV,
and a broad bump centered just below 20 eV.  The sharp peak is
very well reproduced by theory.  It is associated with a collective
resonance of the $\pi$ electrons.  The calculated strength of the
resonance $f$ is =1.1 in the usual sum rule units (i.e. the effective
number of electrons).  The experimental 
strength can be extracted from the data of ref. \cite{hi91}, but the
peak shape has wings that are not simple Lorentzian tails.  Including the 
right-hand wing up to 8.3 eV, but not the left-hand shoulder, the measured
strength is $f=0.9$, which we consider rather good agreement.  
The large broad bump is associated with the $\sigma$ electrons.  We
reproduce its gross structure.  But note that the theoretical distribution
has much finer structure than is observed experimentally.  

\begin{figure}[t!]
\centerline{\epsfig{file=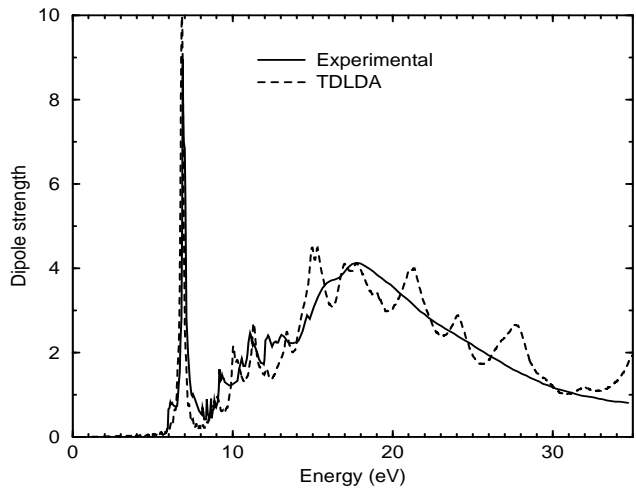,height=7.5cm,width=10cm}}
\caption{In-plane dipole response of benzene molecule, comparing
TDLDA with experiment [15].}
\vspace*{10pt}
\end{figure}

Our next example is C$_{60}$.  Our TDLDA  calculation is
shown in Fig.~4, together with another TDLDA calculation
using the matrix RPA method\cite{ya93}.  In our calculation,
the $\pi$ electron excitation appears
as a single peak at 7 eV, carrying a strength of $f=9$.
Experimentally, the $\pi$ electron transitions are split
into several peaks, with a total strength up
to 7 eV of $f=6$.  There are no absolute measurements of the
strength above 7 eV, but photoionization measurements
exist \cite{he92} that show a broad $\sigma$ resonance,
as in benzene.  The (unnormalized) data is shown in the
figure with diamond symbols. The polarizability of C$_{60}$ has
some interest, because the simplified tight-binding model
gave a number much smaller than experiment, $\alpha\approx 45$ \AA$^3$
compared to the experimental value $\alpha\approx 85 $ \AA$^3$.
The polarizability may be extracted from the dipole strength
function by the integral
\be
\alpha = {e^2\over m} \int {df\over dE} {d E \over E^2}.
\ee
The result is 80 \AA $^3$, in satisfactory agreement with
experiment.

\begin{figure}[t!]
\centerline{\epsfig{file=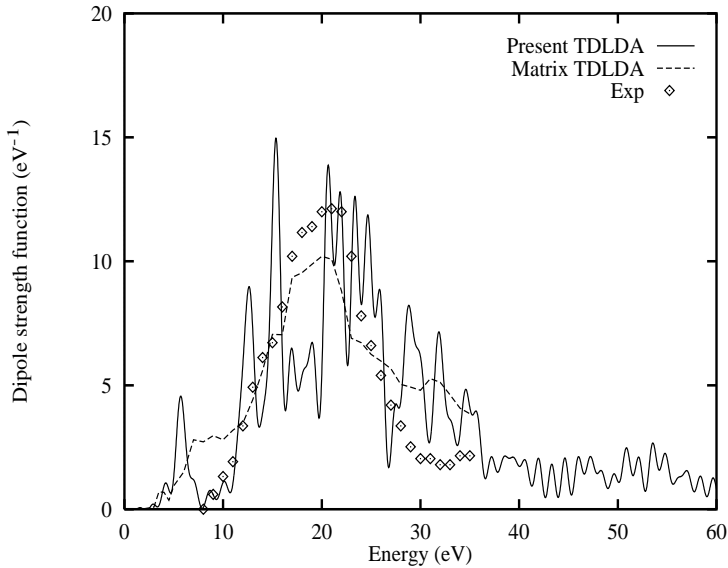,height=7.5cm,width=10cm}}
\caption{Response of C$_{60}$ in TDLDA.  The dashed curve is another
TDLDA calculation, using the particle-hole matrix formulation
[18].}
\vspace*{10pt}
\end{figure}

\subsection{Carbon chains}

The important role of geometry in the dipole response is 
beautifully illustrated by the collective $\pi$
transition in carbon chains and rings.  For these calculations\cite{ya97},
we guessed at the geometry, fixing the nearest neighbor distance at
1.28 \AA~for all cases.  This is the average LDA
equilibrium distance for large rings or chains.  The atoms are in a 
line on the chains and form a circle on the rings. As in the other
examples, there is a strong collective $\pi$ transition
along the axis of the chain or in the plane of the ring.
The energies of the excitations are plotted in
Fig.~5 as a function of the number of atoms. There is
experimental data showing the existence of excitations
at the predicted energies for chains \cite{fo96}.  Unfortunately, the
strengths of the transitions were not measured, so
it is not known whether the observed transitions really 
correspond to the calculated collective excitations.
The stable form of carbon for cluster numbers greater than
ten or so is believed to be the ring configuration, but
no data is available on this form.

\begin{figure}[t!]
\centerline{\epsfig{file=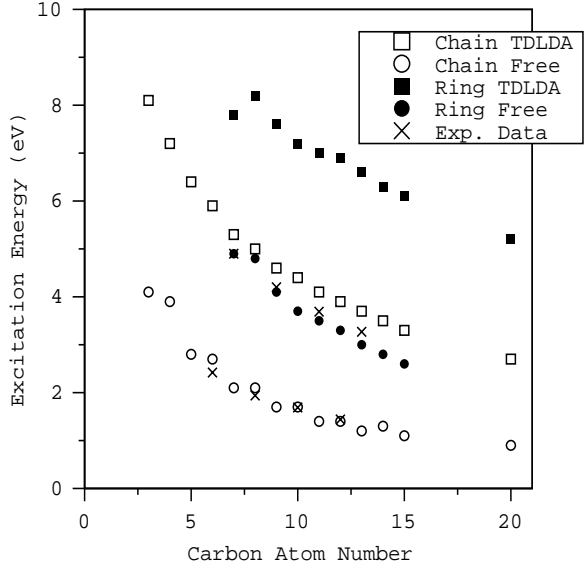,height=7.5cm,width=10cm}}
\caption{Excitation energies of TDLDA collective states and
single-particle excitations in carbon chain and rings.  The
crosses mark experimentally observed excitations [20-22].}
\vspace*{10pt}
\end{figure}

Notice that the excitation energies have a strong but smooth
dependence on the chain length.  The variation is entirely explicable
in terms of simple physics, not requiring at all the
detailed TDLDA calculations.  This is the theme of the
next section of my talk.

\section{Collective energetics in atomic chains}
   The behavior of electromagnetic resonances on infinitely
long wires is known from classical electromagnetic theory.
The dispersion formula for the one-dimensional plasmon
on a long wire reduces to the following expression in the
long-wave length, thin wire limit\cite{go90,li91}. 
\be
\omega^2 = { 4 \pi n_e e^2\over m} q^2 \log{1\over qa}
\ee
where $q$ is the reduced wave number of the plasmon,
$n_e$ is the density of electrons per unit length, and
$a$ is the radius of the wire.  For a finite wire, 
the lowest mode would have a $q$ varying inversely with
the length of the wire $L$.  Thus the lowest mode would
behave as
\be
\omega\sim C {\sqrt{\log(L/a)}\over L}
\ee

This behavior can be extracted from a more quantum approach is
the polarizability estimate of the collective frequency\cite{de93},
\begin{equation}
\label{alpha-estimate}
\omega^2 = {\hbar^2 e^2 N_e \over m  \alpha}
\end{equation}
where $N_e$ is the number of active electrons and $\alpha $ is
the polarizability.  This formula is derived from the ratio of
sum rules, and $N_e$ may be identified with the oscillator strength
$f$ associated with the transition.  For the
linear carbon chains, the transition is associated with $\pi$
electrons and and the number of them in the chain $C_n$ ($n$ odd) is
$$
N_e= 2n-2.
$$
The TDLDA calculations confirm that this is
satisfied at the level of 20\% accuracy.  The polarizability
is harder to estimate.  In ref. \cite{ya97} we model the
polarizability assuming that the chain acts as a perfectly
conducting wire, of some fixed transverse dimension $a$ and
a long length $L$.  Then one can show that for large $L$
the polarizability is given by 
$$
\alpha\approx {L^3\over 24 \log(L)}
$$
The dependence on the number of atoms in the chain follows
form $L\sim n$.
Then inserting the above in eq. (\ref{alpha-estimate}),
we find the large-$n$ behavior
\be
\label{asymp}
\omega \sim {\sqrt{\log n} \over n }.
\ee
This is plotted in Fig.~6 as the solid line, fitted to the
TDLDA result at $n=14$.  We see that the general trend is
reproduced, but the asymptotic behavior is only realized
in long chains.
\begin{figure}[t!]
\centerline{\epsfig{file=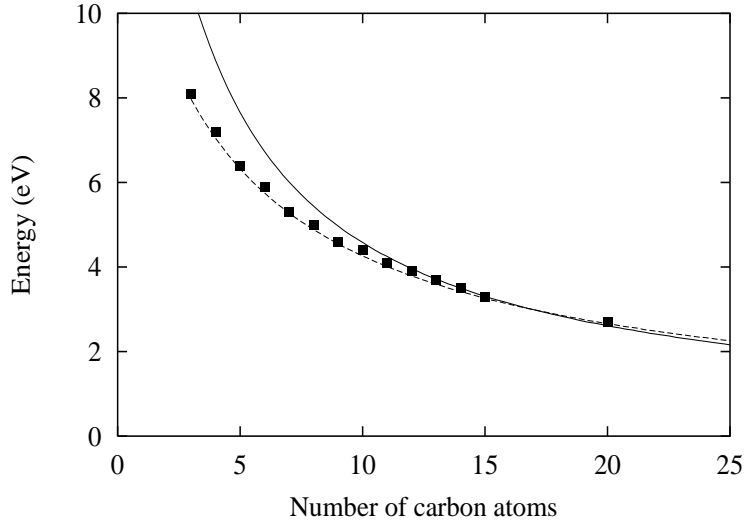,height=7.5cm,width=10cm}}
\caption{Comparison of analytic formulas with the energy
systematics of carbon chains.  Solid line, eq. (10); dashed
line, eq. (11).}
\vspace*{10pt}
\end{figure}

Another classical view is to compare with the polarizability of 
a conducting ellipsoid, which can be calculated 
analytically\cite{bo83}. The formula is
\be
\label{ellipsoid}
\omega^2 = {1-e^2 \over e^2}\Big(-1+{1\over2e}\log{1+e\over
1-e}\Big)\omega_0^2
\ee
where $e$ is related to the ratio of short to long axes,
$R_\perp/R_\parallel$,  
\be
e^2 = 1 - \Big( {R_\perp\over R_{\parallel}}\Big)^2.
\ee
This has been applied to the longitudinal mode in fullerines \cite{br96}.
Its asymptotic behavior is given by eq. (\ref{asymp}) above.  
Remarkably, this formula gives an excellent fit, treating the
two parameters as adjustable.  This is shown as the dashed curve in
Fig.~5.

We finally discuss the relative frequencies of the modes in 
chains and rings.  One expects that for C$_n$ in the form of rings, 
the number of $\pi$ electrons is $2n$, roughly the same as chains.    However, the
polarizability will be quite a bit smaller because of the more
compact geometry.  The ratio of polarizabilities is about a
factor of two, giving a prediction of a 40\% higher collective
resonance.  In fact the TDLDA calculations shows that the
sum rule $f$ is also higher for ring
configurations, and the actual resonance is at about twice the
energy of the chain.  This may be seen in Fig.~5.

\section{What next?}
We have discussed carbon, but from a computational point of
view the next group IV element, silicon, would be very
similar.  The next challenge for numerical TDLDA would be
the group IB metals, the so-called coinage metals.
These have a single valence electron in an $s$ shell, like
the alkali metals, but there is a nearby closed $d$ shell that
cannot be neglected in the response.  Beyond that, there are all 
$d$-shell metals which exhibit broad responses, hardly showing
any trace of the Mie resonance.  But this remains for the
for the future.

\section{Acknowledgment}
We thank R.A. Broglia for calling our attention to eq.~(\ref{ellipsoid}).
This work
is supported by the Department of Energy under
Grant No.~DE-FG06-90ER40561, and by a Grant in Aid
for Scientific Research (No. 08740197) of the Ministry of
Education, Science and Culture (Japan).  Numerical calculations
were performed on the FACOM VPP-500 supercomputer in RIKEN and
the Institute for Solid State Physics, University of Tokyo.


\begin{references}
\bibitem{ya96} K. Yabana and G.F. Bertsch, Phys. Rev. B54 (1996) 4484
\bibitem{fl78} H. Flocard, S. Koonin, and M. Weiss, Phys. Rev. C17 (1978)
1682.
\bibitem{jo89} R.O. Jones and O. Gunnarsson, Rev. Mod. Phys. 61 (1989) 689.
\bibitem{pa88} G. Pacchioni and J. Koutecky, J. Chem. Phys. 88 (1988) 1066.
\bibitem{ko95} M. Kolbuszewski, J. Chem. Phys. 102 (1995) 3679.
\bibitem{br96} H.E. Roman, et al., Chem.Phys. Lett. 251 (1996) 111.
\bibitem{ja96} C. Jamorski, M.E. Casida, D.R. Salahub, J. Chem. Phys.
104 (1996) 5134.
\bibitem{tr91}  N. Troullier and J.L. Martins, Phys. Rev. B43 1993 (1991).
\bibitem{kl82} L. Kleinman and D. Bylander, Phys. Rev. Lett. 48 (1982) 1425.
\bibitem{ek84} W. Ekardt, Phys. Rev. Lett. 52 (1984) 1925;
 W. Ekardt, Phys. Rev. B31 (1985) 6360.
\bibitem{be75} G.F.~Bertsch and S.F.~Tsai, Phys. Reports 18 (1975) 126.
\bibitem{ru96} A. Rubio, et al., Phys. Rev. Lett. 77 (1996) 247;
X. Blase, et al., Phys. Rev. B52 (1995) R2225.
\bibitem{br93} C. Brechignac, et al., Phys. Rev. Lett. 70 (1993) 2036.
\bibitem{ya95} K. Yabana and G.F.~Bertsch, Z. Phys. D32 (1995) 329.
\bibitem{ko72} E.E. Koch and A. Otto, Chem. Phys. Lett. 12 (1972) 476.
\bibitem{hi91} A. Hiraya and K. Shobatake, J. Chem. Phys. 94 (1991) 7700.
\bibitem{he92} I. Hertel, et al., Phys. Rev. Lett. 68 (1992) 784.
\bibitem{ya93} C. Yannouleas, E. Vigezzi, J.M. Pacheco, and R.A. Broglia, Phys. Rev. B47
(1993) 9849; F. Alasia, et al., J. Phys. B27 (1994) L643.
\bibitem{ya97} K. Yabana and G.F.~Bertsch, to be published; xxx.lanl.gov
preprint physics/9612001.
\bibitem{fo96} D. Forney, et al., J. Chem. Phys. 104 (1996) 4954.
\bibitem{fr95} P. Freivogel, et al., J. Chem. Phys. 103 (1995) 54.
\bibitem{fo95} D. Forney, et al., J. Chem. Phys. 103 (1995) 48.
\bibitem{go90} A. Gold and A. Ghazali, Phys. Rev. B41 (1990) 7632, eq.
(23a).
\bibitem{li91} Q.P. Li and S. Das Sarma, Phys. Rev. B43 (1991) 11768, eq.
(2.13).
\bibitem{de93} W. de Heer, Rev. Mod. Phys. 65 (1993) 611.
\bibitem{bo83} C.F. Bohrn and D.R. Huffman,``Absorption and scatttering
of light by small particles", (Wiley, NY, 1983), eq. (5.33).
\end{references}
\end{document}